\documentclass[12pt,epsf,epsfig]{article}
\usepackage{amsmath}

\begin{document}

\begin{center}
{\huge{Dynamical String  Tension  Theories with target space scale invariance SSB and restoration}}  \\
\end{center}

\begin{center}
 E.I. Guendelman  \\
\end{center}

\begin{center}
\ Department of Physics, Ben-Gurion University of the Negev, Beer-Sheva, Israel \\
\end{center}

\begin{center}
\ Frankfurt Institute for Advanced Studies, Giersch Science Center, Campus Riedberg, Frankfurt am Main, Germany \\
\end{center}

\begin{center}
\ Bahamas Advanced Studies Institute and Conferences,  4A Ocean Heights, Hill View Circle, Stella Maris, Long Island, The Bahamas \\
\end{center}
E-mail:  guendel@bgu.ac.il,     

\abstract
The string and brane tensions do not have to be put in by hand, they can be dynamically generated, as in the case when we formulate string and brane theories in the modified measure formalism. Then string and brane tensions appears, but as an additional dynamical degree of freedom . It can be seen however that these string or brane tensions are not universal, but rather each string and each brane generates its own  tension, which can have a different value for each string or brane. The consequences of this for the spectrum of these string and brane  theories is profound both in the ultraviolet behavior as in the low energy physics. There should be also a considerable effect for the effective gravity theories derived from these theories. We consider new background fields that can couple to these new types of extended objects, one of them, the ¨tension scalar¨ is capable of changing locally along the world sheet the value of the tension of the extended object. When many strings probing the same region of space are considered this tension scalar is constrained by the requirement of quantum conformal invariance. For the case of two strings probing the same region of space with different dynamically generated tensions, there are two different metrics, associated to the different strings,  that have to satisfy vacuum Einsteins equations and the consistency of these two Einstein´s equation determines the tension scalar. The universal metric, common to both strings generically does not satisfy Einstein´s equation . The problem is analyzed in the case of a Schwarzschild background and for the cosmological case of a Kasner type solution. In the case of the flat space for the string associated metrics, in the Milne representation, for the case of two types of string tensions, there are solutions with negative string tension at the early universe that  whose tension approaches zero in the late universe and a positive string tension type of strings appears for the late universe with its tension approaching a constant value at the late universe. The universal metric is not flat, instead it represents a non singular bounce cosmology. The case in a warped space time where positive and and negative string tensions are separated by a spontaneously generated wall is also studied, the construction of dynamical tension string theories, where the string tension appears as an integration constant. We also find that the construction of brane world scenarios in the context of these dynamical tension string theories, we discuss avoidance of the Hagedorn temperature possible relaxation of string swampland constraints in dynamical tension string theories, and that the dynamical string theories can bridge between the low and high energy quantum gravity effects, The dynamical modified string theory has target space scale invariance and this target space scale invariance can be restored at the points where the string tension approaches infinity. These models suggest the swampland constraints could be avoided.

\section{Introduction}

String and Brane Theories have been studied as candidates for the theory of all matter and interactions including gravity, in particular string theories \cite{stringtheory}. But string theory has a dimensionful parameter, the tension of the string, in its standard formulation, the same is true for brane theories in their better known formulations.

The appearance of a dimensionful string tension and brane theories from the start appears somewhat unnatural Previously however, in the framework of a Modified Measure Theory, a formalism originally used for gravity theories, see for example \cite{d,b, Hehl, GKatz, DE, MODDM, Cordero, Hidden}, the tension was derived as an additional degree of freedom \cite{a,c,supermod, cnish, T1, T2, T3}. See also the treatment by Townsend and collaborators \cite{xx,xxx}.

A floating cosmological is a generic feature of the modified measure theories of gravity \cite{d,b, Hehl, GKatz, DE, MODDM, Cordero, Hidden}, including the covariant formulation of the unimodular theory   
\cite{HT}, which is in fact a particular case of a modified measure theory, as reviewed in \cite{reviewmodmeas}.

The tension of the string plays a very similar role to the cosmological constant in four dimensional gravity, but the analogous situation and the role of the cosmological constant is quite different to that of the string tension, because while several world sheets of strings
can exist in the same universe and in this way many strings can probe at the same time the same  region of space time,but the same is not the case for the cosmological constant, where every cosmological constant defines necessarily a different universe. 

This paper is organized as follows. After this section, the introduction, in Section 2 we review the modified-measure approach in the string context. In Section 3 we review the modified-measure theory in the brane context. The modified-measure theories of strings and branes rely on two basic elements, the modified measure and the existence of internal gauge fields in the strings and branes, the equations of motion of these gauge fields lead to an equation of motion whose integration constant is the Tension of the extended object.  In Section 4 we discuss the fact that this tension generation (the integration constant) could take place independently for each world sheet separately, which would mean that the string or brane tension is not a fundamental coupling in nature and it could be different for different strings or branes. In section 5  we discuss possible background fields that can be introduced into the theory for the bosonic case and find that  a new field can be introduced that changes the value of the tension of the extended object along its world sheet, we call this the tension scalar .  We present  first the coupling of gauge fields in the extended objects to currents in the world sheet of the extended object that couple to the gauge fields, as a consequence this coupling induces variations of the tension along the world sheet of the extended object. Then we consider a bulk scalar and how this scalar naturally can induce this world sheet current that couples to the internal gauge fields, the equation of motion of the internal gauge field lead to the remarkably simple equation that the local value of the tension along the string is given by $T= g \phi + T _{i} $ , where $g$ is a coupling constant that defines the coupling of the bulk scalar to the world sheet gauge fields and  $ T _{i} $ is an integration constant which can be different for each string in the universe. In section 6 we introduce the target space scale invariance or space time scale symmetry  of the theory, which does not exist in the standard string theory . Then, in section 7,  each string is considered as an independent system that can be quantized. We take into account the string generation by introducing the tension as a function of the scalar field as a factor inside a Polyakov type action with such string tension, then the metric and the factor $g \phi + T _{i} $  enter together in this effective action, so if there was just one string the factor could be incorporated into the metric and the condition of conformal invariance will not say very much about the scalar  $\phi $ , but if many strings are probing the same regions of space time, then considering a background metric $g_{\mu \nu}$ , for each string the ¨string dependent metrics¨  $(\phi + T _{i})g_{\mu \nu}$ appears and in the absence of other background fields, like dilaton and antisymmetric tensor fields, Einstein´s equations apply for each of the metrics $(\phi + T _{i})g_{\mu \nu}$, considering two types of strings with $T _{1 \neq }T _{2}$. We call $g_{\mu \nu}$ the universal metric, which in fact does not necessarily satisfy Einstein´s equations. Section

The problem is analyzed first at a basic level in the case of a Schwarschild backgrounds and for the cosmological case of a Kasner type solution. In the case of the flat space for the string associated metrics, using  the Milne representation for flat space and further considering a $t \rightarrow constant 1/t  $ transformation, we get another conformally related representation of flat space, this leads also to two types of string tensions and for the case of two types of string tensions, we consider solutions with negative string tension at the early universe so that  whose tension approaches zero in the late universe and a positive string tension type of strings appears for the late universe with its tension approaching a constant value at the late universe. The universal metric is not flat, instead it represents a non singular bounce cosmology. The case where positive and and negative string tensions are separated by a spontaneously generated wall is also studied, in warp space times of the type considered by Wesson and collaborators and we discuss these effects concerning the determination of fields that can be introduced into the theory for the bosonic case,  The problem is analyzed in the case of a Schwarschild background and for the cosmological case of a Kasner type solution. In the case of the flat space for the string associated metrics, in the Milne representation, for the case of two types of string tensions, In section 8, we comment that these effects as new types of string interactions which do not exist in the standard string theory, which does not allow for string tension change, also in this same section, we present for our examples preliminary discussed in section 7 and study< in more details where these multi string effects show that  there are solutions containing two metrics related by a conformal transformation which can involve both a negative string tension at the early universe whose tension approaches zero in the late universe and a positive string tension type of strings that appears for the late universe with its tension approaching a constant value at the late universe. The universal metric is not flat, instead it represents a non singular bounce cosmology. The case where positive and negative string tensions are separated by a spontaneously generated wall is also studied. Interesting  warp space times of the type considered by two types of Wesson space times related by a conformal transformation . In section 9 we discuss the application of these ideas for the construction of Braneworld scenarios by considering two flat space times, one in Minkowski coordinates and the other in Minkowski space after a special conformal tranformation, then two surfaces with hyperbolic motion appear where the two strings have tensions which approach infinity, leading to the trapping of the strings inside the two surfaces, which implies the generation of a braneworld scenario. Section 10 discusses the effect of the string tension going to infinite found in previous sections as representing Target space scale invariance restoration. Section 11 discuss the possibility of avoiding of avoiding the  Swampland constraints in the dynamical String tension theory since dynamical string tension also means dynamical Planck scale . Section 12 is an overview where we also add some discussion of avoidance of the Hagedorn temperature as well since it is an effect that is also determined by the string Tension, and if the string tension goes to infinity in some regions of space, that would mean the Hagedorn Temperature would also go to infinity, that is there will be no maximum temperature there . We also discuss in this sections preliminary discussions of results for situations with three different string tensions and hints on what this subject can be handled in future publications.

\section{The Modified Measure Theory String Theory}

The standard world sheet string sigma-model action using a world sheet metric is \cite{pol1}, \cite{pol2}, \cite{pol3}

\begin{equation}\label{eq:1}
S_{sigma-model} = -T\int d^2 \sigma \frac12 \sqrt{-\gamma} \gamma^{ab} \partial_a X^{\mu} \partial_b X^{\nu} g_{\mu \nu}.
\end{equation}

Here $\gamma^{ab}$ is the intrinsic Riemannian metric on the 2-dimensional string worldsheet and $\gamma = det(\gamma_{ab})$; $g_{\mu \nu}$ denotes the Riemannian metric on the embedding spacetime. $T$ is a string tension, a dimension full scale introduced into the theory by hand. \\

From the variations of the action with respect to $\gamma^{ab}$ and $X^{\mu}$ we get the following equations of motion:

\begin{equation} \label{eq:tab}
T_{ab} = (\partial_a X^{\mu} \partial_b X^{\nu} - \frac12 \gamma_{ab}\gamma^{cd}\partial_cX^{\mu}\partial_dX^{\nu}) g_{\mu\nu}=0,
\end{equation}

\begin{equation} \label{eq:3}
\frac{1}{\sqrt{-\gamma}}\partial_a(\sqrt{-\gamma} \gamma^{ab}\partial_b X^{\mu}) + \gamma^{ab} \partial_a X^{\nu} \partial_b X^{\lambda}\Gamma^{\mu}_{\nu\lambda}=0,
\end{equation}

where $\Gamma^{\mu}_{\nu\lambda}$ is the affine connection for the external metric. \\

There are no limitations on employing any other measure of integration different than $\sqrt{-\gamma}$. The only restriction is that it must be a density under arbitrary diffeomorphisms (reparametrizations) on the underlying spacetime manifold. The modified-measure theory is an example of such a theory. \\

In the framework of this theory two additional worldsheet scalar fields $\varphi^i (i=1,2)$ are introduced. A new measure density is

\begin{equation}
\Phi(\varphi) = \frac12 \epsilon_{ij}\epsilon^{ab} \partial_a \varphi^i \partial_b \varphi^j.
\end{equation}

Then the modified bosonic string action is (as formulated first in \cite{a} and latter discussed and generalized also in \cite{c})

\begin{equation} \label{eq:5}
S = -\int d^2 \sigma \Phi(\varphi)(\frac12 \gamma^{ab} \partial_a X^{\mu} \partial_b X^{\nu} g_{\mu\nu} - \frac{\epsilon^{ab}}{2\sqrt{-\gamma}}F_{ab}(A)),
\end{equation}

where $F_{ab}$ is the field-strength  of an auxiliary Abelian gauge field $A_a$: $F_{ab} = \partial_a A_b - \partial_b A_a$. \\

It is important to notice that the action (\ref{eq:5}) is invariant under conformal transformations of the intrinsic measure combined with a diffeomorphism of the measure fields, 

\begin{equation} \label{conformal}
\gamma_{ab} \rightarrow J\gamma_{ab}, 
\end{equation}

\begin{equation} \label{diffeo} 
\varphi^i \rightarrow \varphi^{'i}= \varphi^{'i}(\varphi^i)
\end{equation}
such that 
\begin{equation} \label{measure diffeo} 
\Phi \rightarrow \Phi^{'}= J \Phi
\end{equation}

Here $J$ is the jacobian of the diffeomorphim in the internal measure fields which can be an arbitrary function of the world sheet space time coordinates, so this can called indeed a local conformal symmetry.

To check that the new action is consistent with the sigma-model one, let us derive the equations of motion of the action (\ref{eq:5}). \\

The variation with respect to $\varphi^i$ leads to the following equations of motion:

\begin{equation} \label{eq:6}
\epsilon^{ab} \partial_b \varphi^i \partial_a (\gamma^{cd} \partial_c X^{\mu} \partial_d X^{\nu} g_{\mu\nu} - \frac{\epsilon^{cd}}{\sqrt{-\gamma}}F_{cd}) = 0.
\end{equation}

It implies

\begin{equation} \label{eq:a}
\gamma^{cd} \partial_c X^{\mu} \partial_d X^{\nu} g_{\mu\nu} - \frac{\epsilon^{cd}}{\sqrt{-\gamma}}F_{cd} = M = const.
\end{equation}

The equations of motion with respect to $\gamma^{ab}$ are

\begin{equation} \label{eq:8}
T_{ab} = \partial_a X^{\mu} \partial_b X^{\nu} g_{\mu\nu} - \frac12 \gamma_{ab} \frac{\epsilon^{cd}}{\sqrt{-\gamma}}F_{cd}=0.
\end{equation}

We see that these equations are the same as in the sigma-model formulation (\ref{eq:tab}), (\ref{eq:3}). Namely, taking the trace of (\ref{eq:8}) we get that $M = 0$. By solving $\frac{\epsilon^{cd}}{\sqrt{-\gamma}}F_{cd}$ from (\ref{eq:a}) (with $M = 0$) we obtain (\ref{eq:tab}). \\

A most significant result is obtained by varying the action with respect to $A_a$:

\begin{equation}
\epsilon^{ab} \partial_b (\frac{\Phi(\varphi)}{\sqrt{-\gamma}}) = 0.
\end{equation}

Then by integrating and comparing it with the standard action it is seen that

\begin{equation}
\frac{\Phi(\varphi)}{\sqrt{-\gamma}} = T.
\end{equation}

That is how the string tension $T$ is derived as a world sheet constant of integration opposite to the standard equation (\ref{eq:1}) where the tension is put ad hoc.The variation with respect to $X^{\mu}$ leads to the second sigma-model-type equation (\ref{eq:3}). The idea of modifying the measure of integration proved itself effective and profitable. This can be generalized to incorporate super symmetry, see for example \cite{c}, \cite{cnish}, \cite{supermod} , \cite{T1}.
For other mechanisms for dynamical string tension generation from added string world sheet fields, see for example \cite{xx} and \cite{xxx}. However the fact that this string tension generation is a world sheet effect 
and not a universal uniform string tension generation effect for all strings has not been sufficiently emphasized before. Now we go and review the  Modified Measure Brane Theory
\section{The Modified Measure Brane Theory}

The standard world sheet string sigma-model action using a world sheet metric is \cite{Brane}:

\begin{equation}\label{eq:9}
S_{sigma-model} = -T\int d^d \sigma \frac12 \sqrt{-\gamma}( \gamma^{ab} \partial_a X^{\mu} \partial_b X^{\nu} g_{\mu \nu}+ 2\Lambda).
\end{equation}
Notice that now a cosmological term has been added as well, that was not needed in the usual formulation of the sring theory or in the modified measure formulation of the string theory. In the modified measure formulation of the brane theory such term will not be required. As it is well known, and as we will review, in the standard formulation , this cosmological term needs to be fine tuned.
Here again  $\gamma^{ab}$ is the intrinsic Riemannian metric on the d-dimensional brane worldsheet and $\gamma = det(\gamma_{ab})$; $g_{\mu \nu}$ denotes the Riemannian metric on the embedding spacetime. $T$ is a brane tension, a dimension full scale introduced into the theory by hand. \\

From the variations of the action with respect to $\gamma^{ab}$ and $X^{\mu}$ we get the following equations of motion:

\begin{equation} \label{eq:tab brane}
T_{ab} = (\partial_a X^{\mu} \partial_b X^{\nu} - \frac12 \gamma_{ab}\gamma^{cd}\partial_cX^{\mu}\partial_dX^{\nu}) g_{\mu\nu}-\gamma_{ab}\Lambda =0,
\end{equation}

\begin{equation} \label{eq:Brane}
\frac{1}{\sqrt{-\gamma}}\partial_a(\sqrt{-\gamma} \gamma^{ab}\partial_b X^{\mu}) + \gamma^{ab} \partial_a X^{\nu} \partial_b X^{\lambda}\Gamma^{\mu}_{\nu\lambda}=0,
\end{equation}

where $\Gamma^{\mu}_{\nu\lambda}$ is the affine connection for the external metric. \\

From (\ref{eq:tab brane}) we see that without the cosmological constant term, or if this term is not fine tuned, the equations of motion are inconsistent upon considering the trace for example. This  feature is absent in the modified measure brame theory, where an integation constant, that takes the role of the cosmological term is determined dynamically by the consistency of the equations of motion.

To show  this, again , as in the string case, we consider other measure of integration different than $\sqrt{-\gamma}$. The only restriction is that it must be a density under arbitrary diffeomorphisms (reparametrizations) on the underlying spacetime manifold, as we have seen before while reviewing the modified measure string theory, a metric independent measure build along the lines of what we did in the previous section is an example for that \\

In the framework of  a  $d$ dimensional brane  $d$ additional worldsheet scalar fields $\varphi^i (i=1,2, ...,d)$ are introduced. The new measure density is defined now as

\begin{equation}
\Phi(\varphi) =  \epsilon_{ijk...m}\epsilon^{abc....d} \partial_a \varphi^i \partial_b \varphi^j......\partial_d\varphi^m
\end{equation}

Then the modified bosonic brane action is 

\begin{equation} \label{eq:11}
S = -\int d^d \sigma \Phi(\varphi)(\frac12 \gamma^{ab} \partial_a X^{\mu} \partial_b X^{\nu} g_{\mu\nu} - \frac{\epsilon^{abcd...}}{2\sqrt{-\gamma}}F_{abcd...}(A)),
\end{equation}
where $F_{abcd...}$ is the rank  $d$ totally antisymmetric field-strength  derived from an auxiliary abelian gauge field $A_{acd...}$ of rank  $d-1$: 
$F_{ab....cd} = \partial_a A_{bcd..} +....$ , the totally antisymmetrized curl of a $d-1$ potential $A_{bcd......}$

In the brane modified theory there is a scale invariance, although it is only a global one, as opposed to the string case, where it is local,
\begin{equation} \label{dil}
\gamma_{ab} \rightarrow J\gamma_{ab}, 
\end{equation}

\begin{equation} \label{scalemeasure} 
\varphi^i \rightarrow \varphi^{'i}= \lambda^{ij}(\varphi^j)
\end{equation}
such that 
\begin{equation} \label{measurescale} 
\Phi \rightarrow \Phi^{'}= J \Phi
\end{equation}
That is if $det(\lambda^{ij})= J$, where $J,\lambda^{ij} $ are just a constant number and a constant matrix now, and finally the antisymmetric tensor field $A_{bcd......}$ must transform as
\begin{equation} \label{Ascale}
A_{bcd......} \rightarrow  J^{\frac{d-2}{2}}  A_{bcd......}
\end{equation} 
To check that the new action is consistent with the sigma-model one, let us derive the equations of motion of the action (\ref{eq:9}). 
To start, the variation with respect to $\varphi^i$ leads to the following equations of motion:

\begin{equation} \label{eq:12}
K^{a}_b  \partial_a (\gamma^{cd} \partial_c X^{\mu} \partial_d X^{\nu} g_{\mu\nu} - \frac{\epsilon^{cdef...}}{\sqrt{-\gamma}}F_{cdef...}) = 0.
\end{equation}

Where the matrix $K^{a}_b$ is just like the measure, except that one factor of a derivative of a measure field is missing, so that there are two free indices therefore the determinant of $K^{a}_b$ is a power of the measure, so if the measure is non vanishing, this implies

\begin{equation} \label{eq:a}
\gamma^{cd} \partial_c X^{\mu} \partial_d X^{\nu} g_{\mu\nu} - \frac{\epsilon^{cdef...}}{\sqrt{-\gamma}}F_{cdef...} = M = const.
\end{equation}

The equations of motion with respect to $\gamma^{ab}$ are

\begin{equation} \label{eq:13}
T_{ab} = \partial_a X^{\mu} \partial_b X^{\nu} g_{\mu\nu} - \frac12 \gamma_{ab} \frac{\epsilon^{cdef...}}{\sqrt{-\gamma}}F_{cdef...}=0.
\end{equation}

We will see now that the  final equations  equations are the same as in the sigma-model formulation (\ref{eq:tab}), (\ref{eq:3}). Namely, taking the trace of (\ref{eq:13}),  using also (\ref{eq:a}), and then reinserting into (\ref{eq:13}), we get that $M \neq  0$ now. By solving $\frac{\epsilon^{cdef..}}{\sqrt{-\gamma}}F_{cdef...}$ from (\ref{eq:a})  we obtain a relation between the intrinsic metric and the induced metric if $p \neq 1$, (where $p$ is the number of spacelike world sheet oordinates ($p = d-1$) )  that if we do not deal with the string case  ($p = 1$),\\
\begin{equation}
\gamma_{ab} = \frac{1-p}{M}\partial_a X^{\mu} \partial_b X^{\nu} g_{\mu\nu}
\end{equation}

Using the global scale symmetry of the modified measure brane theory we can set $\frac{1-p}{M} =1$ if $M$ is negative if we wish,  for  $p \neq 1$.  That is, the intrinsic metric in the world sheet becomes equal to the induced metric from the embedding metric, while in the string case there is only a conformal equivalence between the intrinsic metric in the world sheet and the induced metric.
 varying the action with respect to $A_{abc...}$:

\begin{equation}
\epsilon^{abc..d} \partial_d (\frac{\Phi(\varphi)}{\sqrt{-\gamma}}) = 0.
\end{equation}

Then by integrating and comparing it with the standard action it is seen that

\begin{equation}
\frac{\Phi(\varphi)}{\sqrt{-\gamma}} = T.
\end{equation}

That is how the brane tension $T$ is derived as a world sheet constant of integration, opposite to the standard equation (\ref{eq:1}) where the tension is put in from the start. One can see indeed that the string energy excitations will be proportional to $T$ in exactly the same way as it is the case in the standard case. The variation with respect to $X^{\mu}$ leads to the second sigma-model-type equation. The idea of modifying the measure of integration proved itself effective and profitable.

\section{Each String and Each Brane in its own world sheet determines its own  tension. Therefore the  tension is not universal for all strings or branes}

If we look at a single string, the dynamical string tension theories and the standard string  theories appear indeed indistinguishable, there are however more than one string and/or one brane in the universe then, let us now observe indeed that it does not appear that the string tension or the brane tension derived in the sections above correspond to ¨the¨ string or brane tensions of the theory. The derivation of the string or brane tensions in the previous sections holds for a given string or brane, there is no obstacle that for another string or brane these could acquire a different string or brane  tension.  In other words, the string or brane tension is a world sheet constant, but it does not appear to be a universal constant same for all strings and for all branes.
Similar situation takes place in the dynamical string generation proposed by Townsend for example \cite{xx}, in that paper worldsheet fields include an electromagnetic gauge potential. Its equations of motion are those of the Green-Schwarz superstring but with the string tension given by the circulation of the worldsheet electric field around the string. So again ,in \cite{xx} also a string will determine a given tension, but another string may determine another tension. 
If the tension is a universal constant valid for all strings, that would require an explanation in the context of these dynamical tension string theories, for example some kind of interactions that tend to equalize string tensions, or that all  strings in the universe originated from the splittings of one primordial string or some other mechanism. 

In any case, if one believes for example in strings , on the light of the dynamical string tension mechanism being a process that takes place at each string independently, we must ask whether all strings have the same string tension.

\section{Equations for the Background fields and a new background field}
As discussed by Polchinski for example in \cite{Polchinski} , gravity can be introduced in two different ways in string theory. One way is by recognizing the graviton as one of the fundamental excitations of the string, the other is by considering the effective action of the embedding metric, by integrating out the string degrees of freedom and then the embedding metric and other originally external fields acquire dynamics which is enforced by the requirement of a zero beta function.
These equations fortunately appear to be string tension independent for the critical dimension $D=26$ in the bosonic string for example, so they will not be changed by introducing different strings with different string tensions, if these tensions are constant along the the world sheet .

However, in addition to the traditional background fields usually considered in conventional string theory, one may consider as well an additional scalar field that induces currents in the string world sheet and since the current couples to the world sheet gauge fields, this produces a dynamical tension controlled by the external scalar field as shown at the classical level in \cite{Ansoldi}. In the next two subsections we will study how this comes about in two steps, first we introduce world sheet currents that couple to the internal gauge fields in Strings and Branes and second we define a coupling to an external scalar field by defining a world sheet currents that couple to the internal gauge fields in Strings and Branes that is induced by such external scalar field. This is very much in accordance to the philosophy of Schwinger \cite{Schwinger} that proposed long time ago that a field theory must be understood by probing it with external sources. 

As we will see however, there will be a fundamental difference between this background field and the more conventional ones (the metric, the dilaton field and the two index anti symmetric tensor field) which are identified with some string excitations as well. Instead, here we will see that a single string does not provide dynamics for this field, but rather when the condition for world sheet conformal invariance is implemented for two strings which sample the same region of space time, so it represents a collective effect instead. 

\subsection{Introducing world sheet currents that couple to the internal gauge fields in Strings and Branes}

If to the action of the brane (\ref{eq:11}) we add a coupling
to a world-sheet current $j ^{a _{2} \dots{} a _{p+1}}$, $p=d-1$ (the case $p=1$ represents a string) i.e. a term
\begin{equation}
    S _{\mathrm{current}}
    =
    \int d ^{p+1} \sigma
        A _{a _{2} \dots{} a _{p+1}}
        j ^{a _{2} \dots{} a _{p+1}}
    ,
\label{eq:bracuract}
\end{equation}
see \cite{cosmologyandwarped},  \cite{Escaping},  \cite{summary} , \cite{Life},  \cite{Lidhtlikeandbraneworld} for different applications of this.
 Then the variation of the total action with respect to $A _{a _{2} \dots{} a _{p+1}}$
gives
\begin{equation}
    \epsilon ^{a _{1} \dots{} a _{p+1}}
    \partial _{a _{1}}
    \left(
        \frac{\Phi}{\sqrt{- \gamma}}
    \right)
    =
    j ^{a _{2} \dots{} a _{p+1}}
    .
\label{eq:gauvarbracurmodtotact}
\end{equation}
We thus see indeed that, in this case, the dynamical character of the
brane is crucial here.
\subsection{Coupling to a bulk scalar field, the tension field}

Suppose that we have an external scalar field $\phi (x ^{\mu})$
defined in the bulk. From this field we can define the induced
conserved world-sheet current
\begin{equation}
    j ^{a _{1} \dots{} a _{p+1}}
    =
    e \partial _{\mu} \phi
    \frac{\partial X ^{\mu}}{\partial \sigma ^{a}}
    \epsilon ^{a a _{2} \dots a _{p+1}}
    \equiv
    e \partial _{a} \phi
    \epsilon ^{a a _{2} \dots a _{p+1}}
    ,
\label{eq:curfroscafie}
\end{equation}
where $e$ is some coupling constant.

Then (\ref{eq:gauvarbracurmodtotact}) can be integrated to obtain
\begin{equation}
  T =  \frac{\Phi}{\sqrt{- \gamma}}
    =
    e \phi + T _{i}
    ,
\label{eq:solgauvarbracurmodtotact2}
\end{equation}
  The constant of integration $T _{i}$ may vary from one string to the other.
\section{Target space scale invariance and its spontaneous breaking for the Modified Measure Dynamical String Tension Theory} 
Notice that the string theory, has world sheet conformal invariance at the classical level, and this world sheet conformal invariance is requires to be extended to the quantum level.

At the classical level, the ordinary string theory does not have  target space scale invariance, which is very much related to the fact
that there is a definite scale in the theory, the string tension.

Indeed, in the ordinary string theory, a scale transformation of the background metric

$$ g_{\mu \nu}  \rightarrow   \omega g_{\mu \nu}$$
where $\omega $ is a constant, 
is not a symmetry of the Polyakov action, but in the dynamical tension string theory, this transformation is a symmetry provided the world sheet gauge fiends and the  measure transforms as
$$ A_{a}  \rightarrow   \omega A_{a}$$
 $$\Phi(\varphi) \rightarrow   \omega ^{-1} \Phi(\varphi)  $$
 and the tension field transforms in a similar way,
 $$\phi \rightarrow   \omega ^{-1} \phi $$
 As we have seen, the integration of the equations of motions leads to the spontaneous generation of the string tension and at the same time, the spontaneous generation of the target space global scale invariance, since
for this case, (\ref{eq:solgauvarbracurmodtotact2}) is satisfied.
or  equivalently
\begin{equation}
  \Phi
    =
   \sqrt{- \gamma}( e \phi + T _{i})
    ,
\label{eq:solgauvarbracurmodtotact}
\end{equation}

 Notice that the interaction is metric independent. Notice that in the absence of these constants of integration, i.e. , if  $T _{i}= 0$ there is no breaking of the Target space scale invariance, since the measure  and the tension field transform in the same way, but the introduction of non zero constants of integration introduces a spontaneous breaking of the Target space scale invarince. The role of the constants of integrations  $T _{i}$ is analogous to the role of the integrations  $M _{i}$ that we discussed in the context of the gravitational theories.

One may interpret 
(\ref{eq:solgauvarbracurmodtotact} ) as the result of integrating out classically (through integration of equations of motion) or quantum mechanically (by functional integration of the internal gauge field, respecting the boundary condition that characterizes the constant of integration  $T _{i}$ for a given string ). Then replacing 
$ \Phi
    =
   \sqrt{- \gamma}( e \phi + T _{i})$ back into the remaining terms in the action gives a correct effective action for each string. Each string is going to be quantized with each one having a different $ T _{i}$. The consequences of an independent quantization of  many strings with different $ T _{i}$
covering the same region of space time will be studied in the next section.    

A similar exercise can be considered for the Target space scale invariance and its spontaneous breaking for the Modified Measure Dynamical Brane Tension Theory. 

\section{Constraints from quantum conformal invariance on the Tension field, when several strings share the same region of space.}
If we have a scalar field coupled to a string or a brane in the way described in the sub section above, i.e. through the current induced by the scalar field in the extended object,  according to eq. 
(\ref{eq:solgauvarbracurmodtotact}), so we have two sources for the variability of the tension when going from one string to the other: one is the integration constant $T _{i}$ which varies from string to string and the other the local value of the scalar field, which produces also variations of the  tension even within the string or brane world sheet.

As we discussed in the previous section, we can incorporate the result of the tension as a function of scalar field $\phi$, given as $e\phi+T_i$, for a string with the constant of integration $T_i$ by defining the action that produces the correct 
equations of motion for such string, adding also other background fields, the anti symmetric  two index field $A_{\mu \nu}$ that couples to $\epsilon^{ab}\partial_a X^{\mu} \partial_b X^{\nu}$
and the dilaton field $\varphi $ that couples to the topological density $\sqrt{-\gamma} R$
\begin{equation}\label{variablestringtensioneffectiveacton}
S_{i} = -\int d^2 \sigma (e\phi+T_i)\frac12 \sqrt{-\gamma} \gamma^{ab} \partial_a X^{\mu} \partial_b X^{\nu} g_{\mu \nu} + \int d^2 \sigma A_{\mu \nu}\epsilon^{ab}\partial_a X^{\mu} \partial_b X^{\nu}+\int d^2 \sigma \sqrt{-\gamma}\varphi R .
\end{equation}
Notice that if we had just one string, or if all strings will have the same constant of integration $T_i = T_0$.

In any case, it is not our purpose here to do a full generic analysis of all possible background metrics, antisymmetric two index tensor field and dilaton fields, instead, we will take  cases where the dilaton field is a constant or zero, and the antisymmetric two index tensor field is pure gauge or zero, then the demand of conformal invariance for $D=26$ becomes the demand that all the metrics
\begin{equation}\label{tensiondependentmetrics}
g^i_{\mu \nu} =  (e\phi+T_i)g_{\mu \nu}
\end{equation}
will satisfy simultaneously the vacuum Einstein´s equations,

\subsubsection{The case where all all string tensions are the same, i.e.,  $T _{i}=  T _{0}$, and the appearance of a target space conformal invariance }
If  all  $T _{i}=  T _{0}$, we just redefine our background field so that $e\phi+T _{0} \rightarrow  e\phi$ and then in the effective action for all the strings the same combination $e\phi g_{\mu \nu}$, 
and only this combination will be determined by the requirement that the conformal invariance in the world sheet of all strings be preserved quantum mechanically, that is , that the beta function be zero. So in this case we will not be able to determine $e\phi$ and 
$ g_{\mu \nu}$ separately, just the product $e\phi g_{\mu \nu}$,
so the equation obtained from equating the beta function to zero will have the target space conformal invariance 
$e\phi \rightarrow F(x)e\phi $, 
$g_{\mu \nu} \rightarrow F(x)^{-1}g_{\mu \nu} $. 

That is, there is no independent dynamics for the Tension Field in this case.
So in conclusion, if we just look at one string, or if we look at a set of strings, all of them equal string tensions, then the tension field is not observable, can be gauged away by the  target space conformal invariance explained above. Another way to see this is to realize that there is no string excitation of a single string that corresponds to the tension field, unlike the metric, the dilaton or the  two index antisymmetric potential.

On the other hand, if there are at least two types of string tensions, that symmetry will not exist and there is the possibility of determining separately  $e\phi$ and 
$ g_{\mu \nu}$ (setting  the dilaton and the  two index antisymmetric potential to be trivial).

So the tension field dynamics corresponds to a non trivial collective dynamics of strings that cannot be manifest for the equal string tensions case or the single string case. Of course collective effects are very familiar in physics, for example the statistics of fermions and bosons can be understood only when we consider many particles. Here we will see  in the next subsection how the Tension field gets dynamics in the case of two strings with different tensions,   \cite{Escaping},  \cite{summary} , \cite{Life},  \cite{Lidhtlikeandbraneworld}.

\subsubsection{The case where not all string tensions are the same, with special emphasis of two types of strings with   $T _{1} \neq  
T _{2}$,}
The interesting case to consider is therefore many strings with different $T_i$, let us consider the simplest case of two strings, labeled $1$ and $2$ with  $T_1 \neq  T_2$ , then we will have two Einstein´s equations, for $g^1_{\mu \nu} =  (e\phi+T_1)g_{\mu \nu}$ and for  $g^2_{\mu \nu} =  (e\phi+T_2)g_{\mu \nu}$, 

\begin{equation}\label{Einstein1}
R_{\mu \nu} (g^1_{\alpha \beta}) = 0 
\end{equation}
and , at the same time,
\begin{equation}\label{Einstein1}
  R_{\mu \nu} (g^2_{\alpha \beta}) = 0
\end{equation}

These two simultaneous conditions above  impose a constraint on the tension field
 $\phi$, because the metrics $g^1_{\alpha \beta}$ and $g^2_{\alpha \beta}$ are conformally related, but Einstein´s equations are not conformally invariant, so the condition that Einstein´s equations hold  for both  $g^1_{\alpha \beta}$ and $g^2_{\alpha \beta}$
is highly non trivial.

Let us consider the case that one of the metrics, say  $g^2_{\alpha \beta}$ is a Schwarzschild solution, either a 
4 D Schwarzschild solution X a product flat of Torus compactified extra dimensions
\begin{equation}\label{Schwarschild2}
  ds_2^2 = -(1-2GM/r)dt^2 + \frac{dr^2}{1 - 2GM/r} + r^2 d\Omega^2 +dy_4^2 + dy_5^2 +.......+ dy_{25}^2
\end{equation}
or just a 26 D Schwarzschild solution, in this case, it does not appear possible to have a conformally transformed  $g^2_{\alpha \beta}$ for anything else than in the case that the conformal factor that transforms the two metrics is a constant, let us call it c, in that case $g^1_{\alpha \beta}$ is a Schwarzschild solution of the same type, just with a different mass parameter and different sizes of extra dimensions if the compactified solution is considered. Similar consideration holds for the case the 2 metric is a Kasner solution,

\begin{equation}\label{Kasner2}
ds_2^2 = -dt^2 + t^{2p_1}dx_1^2 + t^{2p_2}dx_2^2 +.......+ t^{2p_{25}}dx_{25}^2
\end{equation}
with the well known constraints on the powers $p_i$ that the sum of the $p_i$ s is $1$ and also that  the sum of the squares of the $p_i$ s is also $1$.
Then  in this case also, it does not appear possible to have a conformally transformed  $g^2_{\alpha \beta}$ for anything else than in the case that the conformal factor that transforms the two metrics is a constant, we will find other cases where the conformal factor will not be a constant,  let us call the conformal factor c in general.
One can also study metrics used to describe gravitational radiation,
then again, multiplying by a constant both the background flat space and the perturbation gives us also a solution of vacuum Einstein´s equations. 

Then for these situations, we have,
\begin{equation}\label{relationbetweentensions}
e\phi+T_1 = \Omega^2(e\phi+T_2)
\end{equation}
 which leads to a solution for $e\phi$
 
\begin{equation}\label{solutionforphi}
e\phi  = \frac{\Omega^2 T_2 -T_1}{1-\Omega^2 } 
\end{equation}
which leads to the tensions of the different strings to be
\begin{equation}\label{stringtension1}
 e\phi+T_1 = \frac{\Omega^2(T_2 -T_1)}{1 - \Omega^2} 
\end{equation}
and
  \begin{equation}\label{stringtension2}
 e\phi+T_2 = \frac{(T_2 -T_1)}{1 - \Omega^2} 
\end{equation}

It is important that we were force to consider a multi metric situation. One must also realize that the constant $\Omega^2$ is physical, 
because both metrics live in the same spacetime, so even if $\Omega^2$ is a constant ,  we are not allowed to perform a coordinate transformation, consisting for example of a rescaling of coordinates for one of the metrics and not do the same transformation for the other metric. 

Other way  to see that $\Omega^2$ is physical consist of considering the scalar consisting of the ratio of the two measures $\sqrt{-g^1}$ and $\sqrt{-g^2}$ where $ g^1 = det ( g^1_{\alpha \beta})$ and $ g^2 = det ( g^2_{\alpha \beta})$, and we find that the scalar 
$\frac{\sqrt{-g^1}}{\sqrt{-g^2}} = \Omega^{D}$, showing that $\Omega^2$ is a coordinate invariant. 

Let us study now a case where $\Omega^2$ is not a constant and it may be negative as well, We will also focus on a cosmological case. To find this it is  useful to consider flat space in the Milne representation, $D=4$ this reads,
 \begin{equation}\label{Milne4D}
 ds^2 = -dt^2 + t^{2}(d\chi^2 + sinh^2\chi d\Omega_2^2)
\end{equation}
where $ d\Omega_2^2 $ represent the contribution of the 2 angles to the metric when using spherical coordinates, that is, it represents the metric of a two dimensional sphere of unit radius. In $D$ dimensions we will have a similar expression but now we must introduce the metric of a  $D-2$ unit sphere $ d\Omega_{D-2}^2 $ so we end up with the following metric that we will take as the metric 2
 \begin{equation}\label{MilneD1}
 ds_2^2 = -dt^2 + t^{2}(d\chi^2 + sinh^2\chi d\Omega_{D-2}^2)
\end{equation}

For the metric $1$ we will take the metric that we would obtain from the coordinate transformation $t \rightarrow 1/t $ and we furthermore multiply by a constant $\sigma$, so

 \begin{equation}\label{MilneD2}
 ds_1^2 =\frac{\sigma}{t^4} (-dt^2 + t^{2}(d\chi^2 + sinh^2\chi d\Omega_{D-2}^2))
\end{equation}

Then the equations (\ref{relationbetweentensions}), (\ref{solutionforphi}), (\ref{stringtension1}), (\ref{stringtension2}),
with $\Omega^2= \frac{\sigma}{t^4}$. If we want that the only possible singularities take place at $t=0$ only, we must take $\sigma$ negative,
and therefore the two strings have string tensions of opposite signs, since the factor between the two string tensions is $\Omega^2= \frac{\sigma}{t^4}$ according to (\ref{stringtension1}), (\ref{stringtension2}),  which is negative and therefore now the two strings have opposite tensions.

Whether strings 1 or 2 are the ones with negative tensions depends on the sign of $T_2 -T_1$. If we want the strings with negative tension to exist only in the early universe, we must take  $T_2 -T_1$ to be negative. At the same time there will not be positive tension strings in the early universe, but in the late universe approaches a constant value. The positive string tension are the strings 1, with zero tension in the early universe and the tension $T_1 -T_2$ in the late universe. The negative string tension are the strings 2, with $T_1 -T_2$ tension in the early universe and the tension  zero tension in the late universe. 
 \section{A new type of string interactions for multi string determines string tension dynamics. }
 
 We notice the quantum conformal invariance starts to give useful information concerning the tension field for multi string configurations only after at least two strings with different string tensions covering the same region of space are considered. Then the tension  of one string is correlated with the other, this phenomenon can be characterized as a new type of interaction between strings with different string tensions. This correlation,  achieved through quantum mechanics, can legitimately  called  a new kind of string interaction of a very different nature to those considered in the standard string theory.

 Notice that in the standard string theory two strings with different
 tensions cannot interact, cannot split into two strings with different tensions, etc. . By contrast, in the dynamical string tension theory, these interactions are only triggered when the string tensions are different. Since conventional string interactions cannot change the tension of the strings, since splitting or joining strings does not change the tension, these interaction do not play a role in the dynamics of the tensions of the strings, as opposed to the new interactions that arise at the multi string level considered here.
 
\subsection{The Non Singular Bouncing Solution for the Universal
Metric}
As we have seen when the space time is probed
by two types of strings,
there are two metrics that have to satisfy the vacuum Einstein´s equations, this is enough to solve the problem, The interesting thing however is that the universal metric $ g_{\mu \nu}$
does  not have to satisfy Einstein´s equation. We can see this by solving  $ g_{\mu \nu}$ in terms of one of the metrics , for example from  $g^2_{\mu \nu} =  (e\phi+T_2)g_{\mu \nu}$, we have that

 \begin{equation}\label{universalmetric}
 ds^2 =g_{\mu \nu}dx^{\mu}dx^{\nu} =  (\frac{{1 - \Omega^2}}{T_2 -T_1})(-dt^2 + t^{2}(d\chi^2 + sinh^2\chi d\Omega_{D-2}^2))
\end{equation}
and considering that  $\Omega^2= \frac{\sigma}{t^4} = \frac{-K}{t^4}$ ,
where $K$ is positive. So the coefficient of the hyperbolic D -1 dimensional metric $d\chi^2 + sinh^2\chi d\Omega_{D-2}^2$ is 
$\frac{{t^{2} + \frac{K}{t^2}}}{T_2 -T_1}$, showing a contraction, a bounce and a subsequent expansion. The initial and final spacetimes are flat and satisfy vacuum Einstein´s equations, but not the full space, with most appreciable deviations from   Einstein´s equations at the bouncing time, $t =t*= K^{1/4}$.

One can bring this metric in terms of cosmic time, where the $00$ of  metric component of $g_{\mu \nu}$ is normalized to $1$. One has to notice however that we cannot do the transformation of coordinates only on one of the three metrics we have discussed $g^2_{\mu \nu}, g^1_{\mu \nu}$ and $g_{\mu \nu}$, and if we bring the  the $00$ metric component of the metric $g_{\mu \nu}$ is normalized to $1$, it will not happen simultaneously for  $g^2_{\mu \nu}, g^1_{\mu \nu}$. Having this in mind, the cosmic time coordinate $T$ where  where the $00$ of  metric component of $g_{\mu \nu}$ is normalized to $1$ is defined by 
 \begin{equation}\label{cosmictime}
 dT  = \sqrt{ \frac{{1 + \frac{K}{t^4}}}{T_1 -T_2}}dt 
\end{equation}

So, we see that as $t \rightarrow \infty  $, 
 $T \rightarrow c_1 t$, while for  $t \rightarrow 0 $,  
 $T \rightarrow -c_2/ t$, here $c_1, c_2$ are positive constants. So at large negative cosmic time we have a contacting Milne space, a bounce and the evolution towards an expanding Milne space at large positive cosmic time.
\subsection{ The case where positive and and negative string tensions are separated by a spontaneously generated wall in Wesson warped spaces}

One may wonder if there are similar solutions to the vacuum Einstein´s equations similar  to the Milne space but where instead of time some spacial coordinate would play a similar way. The answer to this question is yes, and these  are the solutions in higher dimensional vacuum General Relativity discovered by Wesson and collaborators, see
\cite{Wesson} and references there. In  five dimensions for example the following warped solution is found,
\begin{equation}\label{Wesson1}
 ds^2 =l^2dt^2 -l^{2} cosh^{2}t (\frac{dr^2}{1-r^{2}} 
 + r^2 d\Omega_{2}^2)) - dl^{2}
\end{equation}
where $l$ is the fourth dimension,
so we see that as in the fourth dimension  $l$  such a solution is homogeneous of degree two, just as the Milne space time was homogeneous of degree two with respect to the time. Notice that maximally symmetric de Sitter space times sub spaces $l =$ constant appear for  instead of euclidean spheres that appear in the Milne Universe for $t =$ constant.

The list of space times of this type is quite large, for example, one cal find solutions of empty GR with Schwarzschild de Sitter subpaces for  $l =$ constant, as in
\begin{equation}\label{Wesson2}
 ds^2 =\frac{\Lambda l^{2}}{3} (dt^2 (1-\frac{2M}{r} -\frac{\Lambda r^2}{3}) - \frac{dr^2}{1-\frac{2M}{r} 
 -\frac{\Lambda r^2}{3}} 
 - r^2 d\Omega_{2}^2) - dl^{2}
\end{equation}
This of course can be extended to $D$ dimensions, where we choose one dimension $l$ to have a factor $l^2$ warp factor for the other dimensions , generically for  $D$ dimensions as in 
\begin{equation}\label{Wessongeneric}
 ds_2^2 =l^{2}\bar{g}_{\mu \nu}(x)dx^{\mu}dx^{\nu} - dl^{2}
\end{equation}
where $\bar{g}_{\mu \nu}(x)$ is a $D-1$ Schwarzschild de Sitter metric for example \cite{Wesson}. This we will take as our $2$ metric, 

In any case, working with this generic  metric of the form  (\ref{Wessongeneric}), but now in $D$ dimensions,  we can perform the inversion transformation $l \rightarrow \frac{1}{l} $, and multiplying also by a factor $ \sigma$ and obtain the conformally transformed metric $1$ that also satisfies the vacuum Einstein´s equations
\begin{equation}\label{Wessongenericinverted}
 ds_1^2 = \sigma l^{-2}\bar{g}_{\mu \nu}(x)dx^{\mu}dx^{\nu} - \sigma\frac{dl^{2}}{l^{4}} = \sigma l^{-4}ds_2^2
\end{equation}

From this point on , the equations the solutions fot the tensions of the $1$ and $2$ strings are the same as in the cosmological case, just that $t  \rightarrow l$, so now $\Omega^2=  \sigma l^{-4}$, so that we now insert this  expression for $\Omega^2$ in (\ref{stringtension1}) and in  
(\ref{stringtension2}), obtaining that on one direction in $l$ negative string tensions dominate, while in the other direction positive string tensions dominate, and we still take $ \sigma = -K$,
where $K$ is positive. \cite{cosmologyandwarped},  \cite{Escaping},  \cite{summary} .

The universal metric, following the steps done for the cosmological case is now ,
\begin{equation}\label{universalmetricwithldependence}
 ds^2 =  (\frac{{1 - \Omega^2}}{T_2 -T_1}) (l^{2}\bar{g}_{\mu \nu}(x)dx^{\mu}dx^{\nu} - dl^{2})
\end{equation}

looking at the coefficient of $\bar{g}_{\mu \nu}(x)dx^{\mu}dx^{\nu}$, the function is $(\frac{{l^{2} + K/l^{2}}}{T_2 -T_1}) $, so the space
time is expanded or contracted as we move in the dimension $l$ by this factor. This factor is minimized at $l* = K^{1/4}$-
We can define a proper length  coordinate $L$ where  where the $ll$ of  metric component of the metric is normalized to $-1$ is defined by 
 \begin{equation}\label{Properlength}
 dL  = \sqrt{ \frac{{1 + \frac{K}{l^4}}}{T_1 -T_2}}dl 
\end{equation}

So, we see that as $l \rightarrow \infty  $, 
 $L \rightarrow c_1 l$, while for  $l \rightarrow 0 $,  
 $L \rightarrow -c_2/ l$, here $c_1, c_2$ are positive constants.
 
Finally we can compare the resulting gravity theories resulting from these multi string effects with known gravity theories discussed in the literature. We have found in the examples discussed so far, the need to define of two conformally related metrics. Since these two are
metrics are conformally related, it is not exactly correct to say that 
we have generated two independent metrics, more accurate is to say that there is a metric say, take the  metric $g^2_{\mu \nu}$, and that there is a measure independent of that metric , $\sqrt {-det({g^1_{\mu \nu})}}$ , so what most resembles this structure are the modified measure theories of gravity  \cite{d,b, Hehl, GKatz, DE, MODDM, Cordero, Hidden}, where there in a metric and a measure that is independent a priori of that metric.

\section{Branewolds from Flat space in Minkowski coordinates and Flat space after a special conformal transformation }

The flat spacetime in Minkowski coordinates is,

 \begin{equation}\label{Minkowski}
 ds_1^2 = \eta_{\alpha \beta} dx^{\alpha} dx^{\beta}
\end{equation}

where $ \eta_{\alpha \beta}$ is the standard Minkowski metric, with 
$ \eta_{00}= 1$, $ \eta_{0i}= 0 $ and $ \eta_{ij}= - \delta_{ij}$.
This is of course a solution of the vacuum Einstein´s equations.

We now consider the conformally transformed metric, \cite{Life},  \cite{Lidhtlikeandbraneworld}, 

 \begin{equation}\label{Conformally transformed Minkowski}
 ds_2^2 = \Omega(x)^2  \eta_{\alpha \beta} dx^{\alpha} dx^{\beta}
\end{equation}
where conformal factor coincides with that obtained from the special conformal transformation
\begin{equation}\label{ special conformal transformation}
x\prime ^{\mu} =  \frac{(x ^{\mu} +a ^{\mu} x^2)}{(1 +2 a_{\nu}x^{\nu} +   a^2 x^2)}
 \end{equation}
for a certain D vector $a_{\nu}$.  which gives $\Omega^2 =\frac{1}{( 1 +2 a_{\mu}x^{\mu} +   a^2 x^2)^2} $
In summary, we have two solutions for the Einstein´s equations,
 $g^1_{\alpha \beta}=\eta_{\alpha \beta}$ and 
 
 \begin{equation}\label{ conformally transformed metric}
 g^2_{\alpha \beta}= \Omega^2\eta_{\alpha \beta} =\frac{1}{( 1 +2 a_{\mu}x^{\mu} +   a^2 x^2)^2} \eta_{\alpha \beta}
 \end{equation}
 
 We can then study the evolution of the tensions using 
 $\Omega^2 =\frac{1}{( 1 +2 a_{\mu}x^{\mu} +  a^2 x^2)^2}$.
 We will consider the cases where  $a^2 \neq 0 $.
    \section*{ The homogeneous and isotropic Universe in Dynamical  String Tension Theories}
  
  We now consider the case when $a^\mu$ is not light like and we will find that for $a^2 \neq 0$, irrespective of sign, i.e. irrespective of whether  $a^\mu$ is space like or time like, we will have thick  Braneworlds  where strings can be constrained  between two concentric spherically symmetric bouncing higher dimensional spheres and where the distance between these two  concentric spherically symmetric bouncing higher dimensional spheres approaches zero at large times.
  The string tensions of the strings one and two are given by
    \begin{equation}\label{stringtension1forBraneworld}
 e\phi+T_1 = \frac{(T_2-T_1)( 1 +2 a_{\mu}x^{\mu} +  a^2 x^2)^2}{( 1 +2 a_{\mu}x^{\mu} +  a^2 x^2)^2-1}=  \frac{(T_2-T_1)( 1 +2 a_{\mu}x^{\mu} +  a^2 x^2)^2}{(2 a_{\mu}x^{\mu} +  a^2 x^2)(2+2 a_{\mu}x^{\mu} +  a^2 x^2)}
\end{equation}
  \begin{equation}\label{stringtension2forBraneworld}
 e\phi+T_2 = \frac{(T_2-T_1)}{( 1 +2 a_{\mu}x^{\mu} +  a^2 x^2)^2-1}=  \frac{(T_2-T_1)}{(2 a_{\mu}x^{\mu} +  a^2 x^2)(2+2 a_{\mu}x^{\mu} +  a^2 x^2)}
\end{equation}
Let us by consider the case where  $a^\mu$ is time like, then without loosing generality we can take  $a^\mu = (A, 0, 0,...,0)$.
Now, in order to get homogeneous and isotropic cosmological solutions
we must  consider the limit $A \rightarrow 0 $ and $(T_2-T_1)\rightarrow 0$, in such a way that $\frac{(T_2-T_1)}{A}= K $, where $K$ is a constant. In that case the spatial dependence in the tensions (\ref{stringtension1forBraneworld})  and (\ref{stringtension2forBraneworld}) drops out and we get, 
 \begin{equation}\label{stringtensionforhomogeneousisotropiccosmology}
 e\phi+T_1 =  e\phi+T_2 = \frac{K}{4t}
 \end{equation}
The embedding metric can now be solved.
\begin{equation}\label{embeddingmetricforhomogeneouscase}
 g_{\mu \nu}  = \frac{1}{(e\phi+T_1)} g^1_{\mu \nu}  =
  \frac{4t}{K}\eta_{\mu \nu}
 \end{equation}
which is not a vacuum metric, as opposed to $\eta_{\mu \nu}$ because of the conformal factor $\frac{4t}{K}$.
 \subsection*{Life of the homogeneous and isotropic Universe and emergence of a Braneworld at large times}
 One should notice that the homogeneous and isotropic solution has been obtained only in the limit $A \rightarrow 0 $ and $(T_2-T_1)\rightarrow 0$, in such a way that $\frac{(T_2-T_1)}{A}= K $, where $K$ is a constant.  If $A $ and $T_2-T_1 $ are small but finite, then for large times, of the order of $1/A$. We can formulate this as an uncertainty principle,
 \begin{equation}\label{uncertainty principle}
 (T_2-T_1)\Delta t \approx constant
 \end{equation}
where we have used that $A$ is of the order of $(T_2-T_1)$.
So a small uncertainty in the tension $(T_2-T_1) $ leads to a long lived homogeneous and isotropic phase, while a big uncertainty in the tension $(T_2-T_1)$ leads to short lived homogeneous and isotropic phase.

  In fact in these situations, for finite $(T_2-T_1) $  and  $A $, it is the case that the string tensions can only change sign by going first to infinity and then come back from minus infinity. We can now recognize at those large times the locations where the string tensions go to infinity, which  are determined by the conditions

\begin{equation}\label{boundariesforBraneworld1}
2 a_{\mu}x^{\mu} +  a^2 x^2 = 0
\end{equation}
or 
\begin{equation}\label{boundariesforBraneworld2}
2 +2 a_{\mu}x^{\mu} +  a^2 x^2 = 0
\end{equation}
Let us start by considering the case where  $a^\mu$ is time like, then without loosing generality we can take  $a^\mu = (A, 0, 0,...,0)$.
In this case the denominator in (\ref{stringtension1forBraneworld}) , (\ref{stringtension2forBraneworld}) is
\begin{equation}\label{denominatortimelike}
(2 a_{\mu}x^{\mu} +  a^2 x^2)(2+2 a_{\mu}x^{\mu} +  a^2 x^2) =
(2At +A^2(t^2-x^2))(2+2At+A^2(t^2-x^2))
\end{equation}

The condition (\ref{boundariesforBraneworld1}), if $A \neq 0$ implies then that
\begin{equation}\label{bubbleboundaryforBraneworld1a}
 x^2_1  + x^2_2 + x^2_3.....+ x^2_{D-1}- (t+ \frac{1}{A})^2 = -\frac{1}{A^2}
\end{equation}
if  $A \rightarrow 0 $, it is more convenient to write this in the form
\begin{equation}\label{bubbleboundaryforBraneworld1aforAgoingtozero}
A( x^2_1  + x^2_2 + x^2_3.....+ x^2_{D-1})- At^2 - 2t = 0
\end{equation}
which for the limit $A \rightarrow 0 $ gives us the single singular point $t=0$, which is the origin of the homogeneous and isotropic cosmological solution.

The other boundary of infinite string tensions is, (\ref{boundariesforBraneworld2}) is given by,
\begin{equation}\label{bubbleboundaryforBraneworld1b}
 x^2_1  + x^2_2 + x^2_3.....+ x^2_{D-1}- (t+ \frac{1}{A})^2 = \frac{1}{A^2}
\end{equation}
This has no limit for $A \rightarrow 0 $, all these points disappear from the physical space (they go to infinity).

For $A \neq 0$ we see that (\ref{bubbleboundaryforBraneworld1b}) represents an exterior boundary which has an bouncing  motion with a minimum radius $\frac{1}{A}$ at $t = - \frac{1}{A}$ , 
The denominator (\ref{denominatortimelike}) is positive between these two bubbles.
So for $T_2 -T_1$ positive the tensions are positive and diverge at the boundaries defined above.

The internal boundary (\ref{bubbleboundaryforBraneworld1a}) exists only for times $t$ smaller than $-\frac{2}{A}$ and bigger than  
$0$, so in the time interval $(-\frac{2}{A},0)$
there is no inner surface of infinite tension strings.
This inner surface collapses to zero radius at  $t=-\frac{2}{A}$
and emerges again from zero radius at $t=0$.

For large positive or negative times, the difference between the upper radius  and the lower radius goes to zero as  $t \rightarrow \infty$

\begin{equation}\label{asymptotic}
\sqrt{\frac{1}{A^2} +(t+ \frac{1}{A})^2 } -\sqrt{-\frac{1}{A^2} +(t+ \frac{1}{A})^2 }\rightarrow \frac{1}{t A^2}\rightarrow 0  
\end{equation}
of course the same holds  $t \rightarrow -\infty$.
This means that for very large early or late times the segment where the strings would be confined (since they will avoid having infinite tension) will be very narrow and the resulting scenario will be that of a brane world for late or early times, while in the bouncing region the inner surface does not exist.
Notice that this kind of brane world scenario is very different to the ones previously studied, in particular both gravity (closed strings) and gauge fields (open strings) are treated on the same footing, since the mechanism that confines the strings between the two surfaces relies only on the string tension becoming very big.

We can ignore the part of the solution where $t<-\frac{2}{A}$ and instead take $t=0$ as the origin of the Universe and only consider positive values of cosmic time because the part of the solution with $t<-\frac{2}{A}$ is disconnected, at least at the classical level from the part of the solution with positive cosmic time.

We see then that for the exact limit of $\Delta T \rightarrow 0$
and $A \rightarrow 0 $ we get a perfect homogeneous and isotropic cosmology, but as $\Delta T$
and $A $ are deformed to be small but finite, the scenario is modified at large  times into a braneworld scenario.

\section{ String Tensions going to infinity as representing Target space scale symmetry restoration, Newton´s  constant as an order parameter} 
In the case of our type of dynamical string tension generation, the generation of a finite sting tension itself . This is because the string tension transforms as the measure under a  Target space scale symmetry, so any finite string generation represents automatically a 
spontaneous symmetry breaking of the Target space scale symmetry .

The case of an infinite string tension is of course different, because multiplying infinity by a finite constant leaves the infinity unchanged, so an infinite string tension represents in this case Target space scale symmetry restoration.

Of course we are more used to have symmetry restoration by defining an order parameter that vanishes. This can be achieved by considering 
the Planck scale. A simple relation between the string tension and Newton´s constant that is  the dimensionful gravitational constant parameter has been found in \cite{Paul} ,

$$\pi T  = (\kappa^2)^{-\frac{1}{d-2}}  $$

and $\kappa^2 = 8\pi G =\frac{1}{M^2_P}  $. 

We see that as the string tension goes to infinity  the Planck scale $M_P$ goes to infinity and the dimensionful gravitational constant parameter goes to zero in what we have argued is the target space scale invariant states. 

On the other hand, the Planck scale and the String tension are proportional, so, for the case of infinite string tension we will have also an infinite Planck scale.
\subsection{ Analogy with the Zee induced gravity model} 
If we compare with gravity theories where the dynamical Planck scale is infinite for the scale symmetric and non zero in a  spontaneously scale symmetric state one can look at the model of Zee \cite{Zee}
\begin{equation}
S = \int d^4 x(\sqrt{-g} (-\frac{\epsilon}{2}\phi^2 R + X-V(\phi)) 
\end{equation}
\label{ZeeModel}

where $$V(\phi) =  \frac{1}{8}\lambda (\phi^2 - v^2)^2 $$ and $X$
being the standard kinetic term for the $\phi$ field.
The constant $v^2$ representing the global scale symmetry breaking parameter,  which determines the Newton constant at low energy (after symmetry breaking) and we avoid choosing  $\epsilon = 1/6$, since we want only global and not local scale invariance for $v^2=0$,

This model can be reformulated as a two measure theory , see  \cite{ZeeasTMT} where the symmetry breaking of scale invariance arises from the integration of the measure fields, which produces  an  integration constant $M$ , corresponding to the constant $v^2$ and which is the analogous of the spontaneous generation of the string tension in the dynamical string tension theory,
Notice that $\phi^2$  is the inverse $16\pi G$, where $G$ is the Newton constant , that is $\phi^2 $  is the dynamical Planck scale squared. As $\phi^2 \rightarrow \infty$ scale symmetry breaking can be ignored, so we identify the very large dynamical Planck scale with the scale symmetry breaking restoration phase.

Planck scale square as order parameter; In the Zee model, we can consider the Planck scale square $\phi^2$ as an  order parameter, $\phi^2$ acquires the vale $v^2$ in the ssb state.  $\phi^2$ going to infinity analog in the dynamical string tension model to  dynamical string tension goes to infinity.
That is, in the dynamical string tension theory it will be the very large string tension sector that will be associated with target space scale symmetry restoration.
Infinite tension as the scale invariant state: The tension, defined as 
$ T =  \frac{\Phi}{\sqrt{- \gamma}} $ does transform under target space scale transformations, since 
 $$\Phi(\varphi) \rightarrow   \omega ^{-1} \Phi(\varphi)  $$
 and $\sqrt{- \gamma}$ does not transform in the Target space scale invariant transformations,  
 so  the tension field transforms in a similar way as $\Phi(\varphi)$ and  $T$ also gets rescaled,
 The integration of the equations of motions of the world sheet gauge fields leads to the spontaneous generation of the string tension, with ssb of target scale symmetry for finite $T$  and  in the case the string tensions become infinite the spontaneous we achieve the restoration of the target space global scale invariance, . So there is no scale invariance for  any state with any finite $T$, that symmetry is spontaneously broken, and  case we have restoration  of the scale symmetry only $T$ is infinity, since then $T$   can be invariant and remain invariant at the same time after being multiplied by a finite constant, since only infinity multiplied by a constant remains the same, i.e. infinity. 
 
 Of course if one is not willing to consider infinities, one may consider the inverse quantity, the Newton constant in the Zee model, which will be zero in the unbroken symmetry phase and acquires a finite value for the ssb phase and in the dynamical string tension theory consider the inverse string tension, that is, what is called the slope defined as $1/(2 \pi T)$, and therefore, zero slope will correspond infinite tension, that corresponds to the unbroken scale invariant phase.
\section{Can Dynamical Tension String Theory recover the Swampland?} 
The standard string theory is argued generates a space of acceptable
theories and a ¨swampland¨
 a space of theories that cannot be correct \cite{Vafa}.

 In a general setting, there are a few statements made where the Planck scale appears \cite{Ooguri}:

 1. Distance conjecture: the statement into the requirement that trans-Planckian excursions can not be allowed for
any fields present in the cosmological evolution.
 
$$ \Delta \phi /M_P < O(1) $$
with $M_P$ being the reduced Planck mass

2,  Due to the difficulties of consistently constructing the meta-stable de-Sitter vacua at the heart
of cosmology it has been further proposed a requirement on possible field potentials of theories in the
Landscape \cite{Ooguri}, given by
either 
$$ M_P  \frac{dV/d\phi}{V} > O(1) $$
or 

$$  - M^2_P  \frac{d^2V/d\phi^2}{V} > O(1) $$

Since these constrains are in some tension with cosmological data, ways to obtain less restrictive versions of the swampland constrains have been also discussed \cite{News}

Of course the dynamical string tension theory provides a very meaningful way to weaken these constraints, since at the points in space time where the string tension goes to infinity, the Planck scale also goes to infinity and the swampland constraints above do not provide any constraint indeed.

Outstanding in this discussion is the possibility of a cosmological constant, or an inflaton potential with a very small slope, so that slow roll, as required by inflation in the early universe, or an adequate description of dark energy in the late universe be possible. In fact for an exponential potential
 $V = \alpha exp(- \lambda \phi)$ , in Planck units $\lambda $ is constrained to be smaller than $\sqrt{2}$.
 This makes cosmology very tough, in particular, negative spatial curvatures seem necessary to achieve only a transient acceleration period \cite{Andriot}, but spatial curvature is not easily compatible with inflation, since zero spatial curvature for the universe is one of the main predictions of inflation!.
 It seems problems like this have been the motivation in \cite{News}
to try to make the swampland less restrictive, but we will choose another avenue to achieve this, the string theory with dynamical tension, in the way we have developed in in this paper.

 In dynamical string tension theories the situation may be better since the Planck scale and the string scale are dynamical so, what should enter in the exponential should be more like something like $V =\alpha exp(-\frac{ \lambda \phi}{M_P})$ , where $M_P$ is the dynamical Planck scale.
 Now, as we have seen, the dynamical  $T$ and therefore also  $M_P$  can become arbitrarily large in some regions of space,  if, as we have seen,  for example for configurations that also eliminate the Hagedorn temperature, so the inflaton potential could be then as flat as needed for an adequate phenomenology. We could then recover the flat potentials discussed in the early part of this paper, or something very similar to it.

 It may seem backward for the traditional string theorists to see somebody try to undue some of the analysis on the swampland, but it may be necessary however to do this, because the swampland constraints are making cosmology impossible or almost impossible for the practical cosmologist. The real Universe appears to be firmly in the Swampland of the conventional string theory. The real Universe favors de Sitter space, flat potentials for inflaton fields, all effects apparently accessible with dynamical tension string theories. Remarkable ideas would have to be discarded away if we do not have an answer to the Swampland conjecture depriving the practical cosmologist of important tools.

\section*{Connecting Modified Measures Theories of Gravity with Modified Measures Theories of Strings, i.e. dynamical String Theories. Dynamical tension strings and dark energy, Bridging high and low energies in quantum gravity } 

We can see that the dynamical tension string theories can be a bridge between  high and low energies in quantum gravity, since a dynamical tension string implies a dynamical Planck scale, and therefore the possibility of bringing the Planck scale to low scales dynamically in some regions and back to high values in some other values, which would imply quantum gravity effects from both low energy ad high energies.

A most important subject was addressed in
our previous section ¨Can Dynamical Tension String Theory recover the Swampland?¨ which  is particularly relevant to the Modified Measures Theories of Gravity, and its possible relation to the dynamical string tension theories, since  the Modified Measures Theories of Gravity,  after the spontaneous breaking of scale invariance predicts inflaton potentials with flat regions, allowing for inflation , early and late dark energies regions, quite useful phenomenological, i.e., an ideal framework of  what we call the ¨practical cosmologist¨ . These constructions are however in the swamp of the standard string theory, but as we argued in the previous section, the tension field dynamically approaching very big values can flatten the inflaton potential.

In our formulation the tension field is governed by the new bulk field, the ¨tension field¨. This tension field should play a very important role in the early inflation and the late Universe, so as to recover the desired flat potentials.

Concerning techniques to study the quantization of the dynamical tension theories, one has to notice that the equation of motion for the above mentioned bulk field, the ¨tension field, is possible in the context of a multi string configuration, which determines such a field, when two strings appear interacting with such a background field. So we are talking about multi string effects, perhaps then string field theory could be of use or some  other formalism that involves more than one string. 

 One should notice that the string theory with strictly one string tension is not enough to account for hadronic phenomena, as pointed out by Andreev \cite{Andreev}, where many string tensions have to be introduced for this purpose, and then the Hagedorn temperature and the Hagedorn phase transition  is eliminated as a consequence. This goes in the same direction in what concerns the swampland conjecture, where we propose the alternative counter conjecture that when many string tensions, like for example in the dynamical string tension theories, the swampland conjecture can be avoided from the modified string theory,

In relation to the difficulties of standard string theory to account for dark energy, if we adopt the swampland conjecture, we notice that the dynamical string tension braneworlds can account for the appearance of a de Sitter space,  this is seen from the hyperbolic motion described in (\ref{bubbleboundaryforBraneworld1b})
when embedded in a higher dimensional space reproduces a de Sitter space, as shown in \cite{BUBBLEFROMFLATSPACES}. In \cite{BUBBLEFROMFLATSPACES} we also show how this bubble universe can be formulated without reference to any string theory of any type, but just by matching two flat spaces, one being Minkowski space and the other being being  Minkowski space time after a special conformal transformation. 

 Likewise, one can explore not stringy analogs of other dynamical string theory solutions, like matching two Wesson type solutions related through an inversion transformation $l \rightarrow constant/l $ through a domain wall, the domain way will be again a true de Sitter space and again we obtain this from the matching of two flat spaces. This will be presented with more details in a future publication.

A historical note concerning how  theories with a dimension full coupling constant have been considered: after the development of QED and latter of the standard model which involved ekectro weak and strong interactions which was all based on applying the gauge principle, characterized by  dimensionless gauge couplings and other couplings with dimensions of positive powers of mass, but gravity did not follow this pattern. Therefore it is not such a surprise that string theory with a string tension that manifestly violates the Target space scale invariance in any case accepted. Here however we want to go back and insist on a  Target space scale invariance that can be spontaneously broken, but it can be also restored in some regions or phases of space time, resulting in effects not found in the more conventional string theories.

 The dynamical sting tension theory provides an ideal tool to bridge between high and low energies in quantum gravity, this is because the string tension determines the Planck scale and therefore, locally the Plank scale could change and therefore the scale of quantum gravity effects could be either high at some points in space time or low at some other points of space time. This will be a subject to be developed with my colleagues of the  COST ACTION CA23130 - Bridging high and low energies in search of quantum gravity (BridgeQG).

 \section{Dynamically Generated String and Brane Tension Theories for Cosmology, Warped Space times and Braneworlds and target space scale invariance restoration. Higher multistring tensions}
 
 \subsection{Review of Results in this paper.}The consideration of dynamical tension generation lead us to theories where each extended object, either strings or branes generates dynamically its own string tension. This has profound consequences. 
 We in particular discuss a new background scalar field, the ¨tension¨ field that can change dynamically the value of the tension along the world sheet of the extended objects. This new field introduced in the theory of extended objects appears to have very important consequences in various cosmological and other scenarios.
 
We have studied in particular how from the integration of the world sheet gauge fields, that control the dynamics of the dynamical tension, that the tension in a given string labeled by $i$ is given by $e\phi+T_i$, where $T_i$ is a constant of integration that may differ for different extended objects. 

Then we consider the constraint from the quantum conformal invariance in a situation where at least two strings with constants of integration  $T_1$ and $T_2$, with $T_1 \neq T_2$ are considered and both strings probe the same region of spacetime. For simplicity we consider cases where only the background metric is non trivial, while the dilaton is taken as a constant and the anti symmetric tensor field is zero or pure gauge.

We construct actions that reproduce the equations with tensions 
$e\phi+T_1$ and $e\phi+T_2$, then the requirement of quantum conformal invariance, independently for the string 1 or the string 2 are that 
the auxiliary metrics $g^1_{\mu \nu} =  (e\phi+T_1)g_{\mu \nu}$ and $g^2_{\mu \nu} =  (e\phi+T_2)g_{\mu \nu}$ both satisfy the vacuum Einstein´s equations . The metric $g_{\mu \nu}$ we call the Universal metric. These three metrics,  $g^1_{\mu \nu}, g_{\mu \nu}$ and  $g^2_{\mu \nu} $ are conformally related, the conformal relation between  $g^1_{\mu \nu} =  (e\phi+T_1)g_{\mu \nu}$ anf  $g^2_{\mu \nu} =  (e\phi+T_2)g_{\mu \nu}$, implies  $  (e\phi+T_1) = Omega^2 (e\phi+T_2)$ 
where $\Omega^2\neq 1 $ if  $T_1 \neq T_2$ . $\Omega^2$ could be a constant.
There are many ways to satisfy this, for example scaling a Schwarzschild metric gives us again a Schwarzschild metric, scaling a Kasner solution gives us again a Kasner solution. Notice that $c$ is physical, since any coordinate transformation has to be done on the two metrics, but then the ratio of the densities  
$\frac{\sqrt{-g^1}}{\sqrt{-g^2}} = \Omega^{D}$, is a scalar, showing that c is a coordinate invariant. In this case when expressing the equations of motion of the two strings with the same coordinates, if $\Omega^2$ is a constant, they both satisfy the sage geodesic equation (the factor $\Omega^2$ comes out of the action and does not affect the geodesic equation).

We have obtained the case $\Omega^2$ not a constant in two types of situations, 
one considering the Milne Universe in $D$ dimensions for the $2$ metric and for the $1$ metric, the one that is obtained by the transformaion $t \rightarrow  1/t$  and then multiplying by a constant
$\sigma$, so $\Omega^2= \sigma/t^4$, This allows us to solve for the tensions of the two types of strings. At the early Universe we have negative string tensions, which then go to zero tension and at the same time the positive string tensions appear and those positive string tensions become constant in the late universe. The universal metric is non singular and experiences a bounce. Negative and positive string tensions appear separated by time.

Considering warped  Universes of the Wesson type in $D$ dimensions for the $2$ metric and for the $1$ metric, the one that is obtained by the transformaion $l \rightarrow  1/l$  and then multiplying by a constant
$\sigma$, so $\Omega^2= \sigma/l^4$, This allows us to solve for the tensions of the two types of strings. On one direction of the warp coordinate  we have negative string tensions, which then go to zero tension and at the same time the positive string tensions appear in the other direction of the warp coordinate and those positive string tensions become constant tension as get deep far in that direction. The universal metric is non singular and experiences a wormhole behavior. Negative and positive string tensions appear separated by spacial direction. 

  While the 1 and 2 metrics are the metrics with which the strings 1 and 2 follow geodesics (which are the same equations if the conformal factor between these two metrics is a constant) and for $g_{\mu \nu} $, neither strings $1$ or $2$ follow geodesics with respect to this metric, unless  the conformal factor between  metrics 1 and 2 is a constant, but a bound state of strings 1 and 2 may follow geodesics with respect to this metric. Whether such bound states exist and whether such conjecture is true is an open question at this point.
The universal metric $g_{\mu \nu} $ does not satisfy Einstein´s equations.

Notice also that the warped Wesson type space times are cosmological space times, since the warp de Sitter or Schwarzschild de Sitter spaces. So they may provide a different perspective on the cosmological constant problem , as Wesson suggested. 

One very interesting problem could be to consider the effect of the warp of the spacetime, as it has been done in more conventional scenarios in \cite{warpgravitywaves}

Although our main emphasis here has been on strings, many formulas have been given for the dynamical tension brane theory as well, but then quantum considerations for the case of branes are not so  reliable, nevertheless some progress in this direction should be explored also.

We have in this paper started the study of the effective  gravitational solutions  generated by these dynamical string theories with many strings probing the same region of space. The analysis here has been concentrated on simple special cases, clearly a more generic approach is necessary, in particular one should study solutions with non trivial dilaton and antisymmetric tensor fields, as well as cosmological and warped solutions with all kind of matter.

Comparing the resulting gravity theories resulting from these multi string effects with known gravity theories discussed in the literature. We have found, in the examples discussed so far, the need to define of two conformally related metrics. Since these two are
metrics are conformally related, it is not exactly correct to say that 
we have generated two independent metrics, more accurate is to say that there is a metric say, take the  metric $g^2_{\mu \nu}$, and that there is a measure independent of that metric , $\sqrt {-det({g^1_{\mu \nu})}}$ , so what most resembles this structure are the modified measure theories of gravity  \cite{d,b, Hehl, GKatz, DE, MODDM, Cordero, Hidden}, where there in a metric and a measure that is independent a priori of that metric.
The tension of the strings enters not only in the effective gravity the string theory generates, but also on the spectrum of the strings themselves, these aspects have to be further investigated. 

We can see also that in the case that both the metric $1$ and the metric $2$ are two flat metrics, like in our studies of cosmological solutions, or for the warped spaces in the case the warped space  \ref{Wessongeneric} is defined with  $\bar{g}_{\mu \nu}(x)$ being de Sitter and not the more generic Schwarzschild de Sitter case, the solutions are most likely exact and not just to first order in the slope, since not only the Ricci tensor vanish but also the Riemann curvature and so will all higher curvature corrections to the beta function. 
In other paper we will address other scenarios for cosmology and warped spaces where we will emphasize the effect of dynamical string tensions on the possibility of escaping the Hagedorn temperature, this could be possible if the string tensions become very large at the early Universe or at some position in the warped coordinate  \cite{noHagedorn}. In this case in certain regions of space both type of string tensions are positive and go to infinity in sch a way that their ratio is one and for equal tension strings the more conventional string interactions (besides the one governed by the tension field), have a chance to operate. Notice that a multi-string tension scenario formulated phenomenologically in order to avoid the Hagedorn temperature was formulated by Oleg Andreev \cite{Andreev}, which seems to be the only other place in the literature, in addition to our research, where a multi string tension scenario has been discussed.

We have seen the possibility of obtaining brane-world solutions, where the string tensions become infinity between two hyperbolic expanding surfaces, where the strings must be confined, signaling the strings must be confined between these two surfaces and showing therefore the existence of a new type of brane-world scenario.

As we pointed out, in some of the examples above we have regions of spacetime where tensions of the strings go to infinity. This in turn opens the possibility of the disappearance of the Hagedorn temperature and for the re-apparition of the swampland (i.e. avoidance of the swampland constraints) as defined by the normal string theory, since where the dynamical tensions become infinite so does the dynamical Planck scale and therefore the swampland constraints can be avoided in those regions of space-time.
 \subsection{ Higher multistring tensions case.}
Finally, we have to observe that we have only studied here the first possible effect from the multi-string tension theory, working with only two different string tensions. One could ask what if we have three or more different string tensions present in the same of space time?
Well in this case if the tensions are not all constants, it seems we will not be able to use the simplification where we consider the dilaton and the antisymmetric two index tensor field trivial; we would need now for the case of three different string tensions more degrees of freedom for a consistent solution, which would mean the dilaton and the antisymmetric two index tensor field will be non trivial. 

Let us study some preliminary results for three strings with different string tensions, generalizing  (\ref{relationbetweentensions}), (\ref{solutionforphi}), (\ref{stringtension1}) and (\ref{stringtension2}).
Then for these situations, we have, considering $\Omega_{12}^2$ as the conformal factor between string metric 1 and string metric 2, and therefore we have the relation between the string tensions 1 and 2, 
\begin{equation}\label{relationbetweentensions12}
e\phi+T_1 = \Omega_{12}^2(e\phi+T_2)
\end{equation}
 which leads to a solution for $e\phi$
 
\begin{equation}\label{solutionforphi12}
e\phi  = \frac{\Omega_{12}^2 T_2 -T_1}{1-\Omega_{12}^2 } 
\end{equation}
which leads to the tensions of the different strings 1 and 2 to be
\begin{equation}\label{stringtension1 12}
 e\phi+T_1 = \frac{\Omega_{12}^2(T_2 -T_1)}{1 - \Omega_{12}^2} 
\end{equation}
and
  \begin{equation}\label{stringtension2 12}
 e\phi+T_2 = \frac{(T_2 -T_1)}{1 - \Omega_{12}^2} 
\end{equation}
But now we have the string 3 and its relation to string 2. 
considering $\Omega_{23}^2$ as the conformal factor between string metric 2 and string metric 3, and therefore we have the relation between the string tensions 2 and 3,

\begin{equation}\label{relationbetweentensions23}
e\phi+T_2 = \Omega_{23}^2(e\phi+T_3)
\end{equation}
 which leads to another solution for $e\phi$
 \begin{equation}\label{solutionforphi23}
e\phi  = \frac{\Omega_{23}^2 T_3 -T_2}{1-\Omega_{23}^2 } 
\end{equation}
Now  we have to decide what is our input and what is our output. There are many posibilities, The most simple case is to choose $\Omega_{12}^2$ to be a constant, like in the transformation between two Schwarzchild spaces or in the transformation between Kasner spaces, all of them satisfying Vacuum Einstein equations, then for the third metric we solve $\Omega_{23}^2 $ from equating the two expressions for $ e\phi$ , equations  (\ref{solutionforphi12}) and
(\ref{solutionforphi23}), which gives
\begin{equation}\label{solutionforOMEGA23}
\Omega_{23} = \frac{ T_1+ \Omega_{12}^2( T_3 - T_2)}{\Omega_{12}^2(T_3 - T_2) - T_3 } 
\end{equation}
also a constant value for $\Omega_{23}^2$, in this case all the three string metrics will satisfy vacuum Einstein equations and the matter is closed.

Things become more complicated if we want to consider a non-constant $\Omega_{23}^2$, like the one obtained from a special conformal transformation, since then the solution for $\Omega_{23}^2$ from \ref{solutionforOMEGA23} is not anymore a special conformal transformation, so the the third string metric will not satisfy the vacuum Einstein equation, 
showing that for consistent non constant conformal we hace to activate the dilaton and the two index antisymmetic tensor fiels. Solutions along these lines will be studied in a future research, without assuming from the start any of the conformal factors that relate the different string metrics, that was possible in the two string tension case,

 \section*{Acknowledgements} 
I am  grateful to  COSMOVERSE,  COST ACTION CA21136 and  to COST ACTION CA23130 - Bridging high and low energies in search of quantum gravity (BridgeQG),  to Ben-Gurion University of the Negev for generous support. I also want to thank Professor Juan Maldacena for pointing out that standard string interactions do not work for strings with different tensions a point that was incorporated in the paper, to Professor Paul Steinhardt for discussions on dynamical string tension theories and Target space scale invariance, to  Oleg Andreev for discussion on multi  tension theories , David Andriot for discussions on the string  swampland  and my collaborators in diverse additional aspects reviewed here, Emil Nissimov, Svetlana Pacheva, Douglas Singleton, Sergio del Campo (RIP) , Alexander Kaganovich, Ramon Herrera, Pedro Labraña, Subhash Rajpoot,  Hitoshi Nishino (RIP), Euro Spallucci,  Stefano Ansoldi,
Claudio Paganini,  Felix Finster,  and to  Yoelsy Leyva and Giovanni Ottarola for conversations and for their invitation to talk on this subject in the  International Joint Meeting on Cosmology and Gravitation, IJMCG 2024, https://ijmcg-uta.cl/home/
Thanking them and the other local organizers of this conference from the University of Tarapaca for their hospitality in Arica, Chile.

\end{document}